\documentclass[aps,pre,12pt]{revtex4-1}
\usepackage{graphicx}
\usepackage{amsmath}
\usepackage{graphicx}
\usepackage{epstopdf}
\usepackage{amsmath}
\usepackage{amssymb}
\usepackage{textcomp}
\usepackage{default}
\usepackage{default}
\newcommand{\be}{\begin{equation}}
\newcommand{\bea}{\begin{eqnarray}}
\newcommand{\bc}{\begin{center}}            
\newcommand{\ee}{\end{equation}}
\newcommand{\eea}{\end{eqnarray}}
\newcommand{\ec}{\end{center}}
\newcommand{\baa}{\begin{eqnarray*}}
\newcommand{\eaa}{\end{eqnarray*}}
\begin{document}
\title{Pythagorean means and Carnot machines:\\
When music meets heat}
\author{Ramandeep S. Johal}
\email{rsjohal@iisermohali.ac.in}
\affiliation{ Department of Physical Sciences, \\ 
Indian Institute of Science Education and Research Mohali, 
Sector 81, S.A.S. Nagar, Manauli PO 140306, Punjab, India}
\begin{abstract}
Some interesting relations between Pythagorean means (arithmetic, geometric
and harmonic means) and the coefficients of performance of reversible 
Carnot machines (heat engine, refrigerator and heat pump) are discussed.
\end{abstract}
\maketitle
\section{ Introduction}
Mean is an important concept in mathematics and statistics.
We will consider the Pythagorean means, defined as follows:
\bea 
A(x_1, x_2,\cdots, x_n) & = & \frac{1}{n} \sum_{i=1}^{n} x_i, \\
G(x_1, x_2,\cdots, x_n) & = &  \prod_{i=1}^{n} x_{i}^{1/n}, \\
H(x_1, x_2,\cdots, x_n) & = & n \left( \sum_{i=1}^{n} x_{i}^{-1}
\right)^{-1}
\eea
which denote the arithmetic mean, the geometric mean, and the harmonic mean,
respectively. Here, $\{x_i| i=1,2,...,n \}$ represents a set of positive, real 
numbers. Historically, these means are attributed to the ancient Greek mathematician
Pythagorus, who found their use to represent ratios on a musical scale  (see Box 1).  
There are some common properties satisfied by the means 
${M}(x_1,...,x_n)$:

(i) ${\rm min}(x_1,...,x_n) \leqslant  {M}(x_1,...,x_n)
\leqslant {\rm max}(x_1,...,x_n)$.
So,  ${M}(x_1,...,x_n) = {M}(x,...,x) =x$.

(ii) A mean is homogeneous function of degree one:
${M}(\lambda x_1,...,\lambda x_n) = \lambda {M}(x_1,...,x_n)$, 
for all permissible real $\lambda$. 

(iii) A mean is called {\it symmetric}, if it is invariant under permutations of
the indices. Thus 
${M}(x_1,...,x_n) = {M}(x_{\sigma(1)},...,x_{\sigma(n)})$.

Now, for a given set of numbers, when at least two numbers are distinct,
there exist elementary inequalities between
the means, for example:  $ { H} \leqslant {G} \leqslant {A}$.
To give commonplace examples of the occurence of Pythagorean means, 
suppose that you complete two legs of a journey, spending equal time $t$ in
each of them, and traveling with average speeds of $v_1$ and $v_2$,
then your average speed for the total trip is equal to $H(v_1,v_2)$.
On the other hand, if you cover equal distances in the two legs
(e.g. for a return journey),
at average speeds of $v_1$ and $v_2$, then the average speed 
for the total trip would be $A(v_1,v_2)$. On the other hand,
the geometric mean 
may be seen as the compound average growth rate, 
$G(r_1,r_2) = \sqrt{r_1 r_2}$, over a principal amount 
which gets the growth rates of $r_1$ and $r_2$,
in the first and the second years, respectively. 
The inequalities between various means are useful in 
the proofs of various physical laws. For applications
in thermodynamics, see Box 2.

In this paper, we shall observe 
a natural instance of Pythagorean means in the analysis 
of coefficients of performance (COP) of ideal thermodynamic machines,
in particular, heat engines, refrigerators and heat pumps.
It is assumed that we are provided with heat reservoirs at various temperatures 
representing the set of real, positive numbers over which 
these means may be defined, and we can run these machines 
by coupling them to the reservoirs. 
 
\section{Performance measures for Carnot machines}
We shall first discuss the case of Pythagorean means for two 
numbers in the context of heat engines, refrigerators and heat pumps.
We start by defining the COPs for these machines.

A {\it heat engine} is a device that converts heat into work, as the heat flows
from a reservoir at a constant hot temperature $T_h$ to one at a cold
temperature $T_c$. The COP for an engine is called its efficiency ($E$)
which is defined as the ratio of the work generated ($W$)
to the heat absorbed ($Q_h$) from the hot reservoir, $E = W/Q_h$. For an ideal 
or reversible heat engine which does not increase entropy
of the universe, the efficiency is maximal called Carnot efficiency,
which depends only on the ratio of the two temperatures:
\be
E_{hc} = \frac{T_h - T_c}{T_h}.
\label{eff}
\ee
A {\it refrigerator} is a device that helps to cool a system 
at temperature $T_c$, by extracting 
heat from it ($Q_c$) and dumping this heat into a warm environment at $T_h$. Due
to the constraint of the second law of thermodynamics,
this process cannot proceed spontaneously,
and so external work ($W$) has to be spent to transfer the heat.
The COP of refrigerator is defined as $R = Q_c/W$, which for an 
ideal refrigerator is given by   
\be
R_{hc} = \frac{T_c}{T_h-T_c}. 
\label{cxir} 
\ee
A refrigerator with an irreversible operation 
has a lower COP, implying that it requires a larger
amount of work to extract the same amount of heat 
from the cold system, as compared to the reversible case. 

A {\it heat pump} also transfers heat from a cold to hot system 
at the expense of external work, 
but in this case we are interested to warm the system
(such as interiors of a building in winter) 
by extracting heat from the cold environment. 
So the COP of a pump is defined as the heat dumped into the warm system 
($Q_h$) per unit work expended $W$ for the purpose. Again, for an  
ideal pump, the COP is given by
\be
P_{hc} = \frac{T_h}{T_h-T_c},
\label{cxip}
\ee
where $T_h$ is the temperature of the interiors, and $T_c$ is the ambient
temperature. Note that, by definition, for finite temperatures, 
$0<E<1$, $R>0$ and $P >1$. 

Let us now consider another reservoir (warm) 
with temperature labeled as $T_i$ such that 
$T_h > T_i > T_c$. Suppose, we are allowed to run a Carnot engine 
between reservoirs at $T_h$ and $T_i$, or, between $T_i$ and $T_c$.
The respective efficiencies are : $E_{hi} = (T_h - T_i)/T_h$
and $E_{ic} = (T_i - T_c)/T_i$. We can ask: for what value  
of $T_i$, we have the situation: $E_{hi} = E_{ic}$? It follows 
that at $T_i = \sqrt{T_h T_c}$, the two efficiencies are equal. 
Interestingly, the answer is invariant if we replace 
both the engines by two Carnot refrigerators or pumps, and require 
the equality of their respective COPs.
In other words, if we set $R_{hi} = R_{ic}$, or $P_{hi} = P_{ic}$,
the solution $T_i$ is equal to the geometric mean temperature
of $T_h$ and $T_c$. We address these cases as $EE$, $RR$ and $PP$,
respectively.

Further, we may consider a refrigerator between $T_h$ and $T_i$, 
and a pump between $T_i$ and $T_c$, and require that $R_{hi} = P_{ic}$.
This condition yields, $T_i = (T_h + T_c)/2 = A(T_h,T_c)$. 
Conversely, for the situation when $P_{hi} = R_{ic}$,
we get $T_i = H(T_h, T_c)$. The latter cases are addressed as $RP$ and $PR$,
respectively. Thus, we note an interesting
instance of Pythagorean means in the context of Carnot machines.
These cases are depicted in Fig. 1. See also Table 1.

\section{Generalizations}
Consider a set of $k$ heat reservoirs with temperatures 
ordered as follows:
\be
T_1 > T_2 > \cdots > T_{k-1} > T_k.
\label{t1k}
\ee
Suppose we run a Carnot engine between a pair of reservoirs,  
one being a hot reservoir at $\{T_i | i=1,2,...,k-1 \}$ and the
other being the coldest
reservoir within the set, at $T_k$. The efficiency of a particular
engine would be $E_{ik} = (T_i - T_k)/T_i$. 
Then the average efficiency of all these engines 
may be defined as the arithmetic
mean of the individual efficiencies,
$\bar{E} = \sum_i E_{ik} /(k-1)$. It is easy
to verify that 
\be
\bar{E} = \frac{\bar{T} - T_k}{\bar{T}}
\ee
where $\bar{T} = H(T_1, T_2,...,T_{k-1})$. 
Thus one can state this result as:

$(a)$ {\it The arithmetic mean
of the efficiencies $\{E_{ik} \}$
between $\{T_i\}$ and $T_k$, where $i=1,2,....,k-1$,
is equal to the efficiency of
a Carnot engine between $\bar{T}$ and $T_k$, 
where $\bar{T}$ is harmonic mean
of the $\{T_i\}$.}

We can make similar statements, when the machines in 
case are refrigerators or pumps, connecting 
a certain mean COP of the machine, to 
a certain mean over the hot reservoir temperatures.
These statements are given below. The said connection
is also shown in Table II.

$(b)$ {\it The harmonic 
mean of $\{R_{ik}\}$ between $T_i$ and $T_k$, where $i=1,2,....,k-1$
is equal to the $R$-coefficient of a refrigerator between $T_A$ and $T_k$,
where $T_A$ is the arithmetic mean of $\{T_i\}$.}

{\bf Proof:}
We can write 
\be 
\sum_{i=1}^{k-1} {R_{ik}^{-1}} = \sum_{i=1}^{k-1} \frac{T_i-T_k}{T_k}.
\label{sumr}
\ee
which can be expressed as:
\be
\frac{1}{k-1} \sum_{i=1}^{k-1} {R_{ik}^{-1}} = 
\frac{\sum_i T_i /(k-1)-T_k}{T_k}
\label{sumrh}
\ee
or, 
\be 
R_{H}^{-1} = \frac{T_A - T_k}{T_k},
\label{rh}
\ee
where, $R_H$ is the harmonic mean of $\{R_{ik}\}$, and 
rhs is the inverse of the  
$R$-coefficient ($R_{Ak}$) between $T_A$  and $T_k$.

$(c)$ {\it The harmonic mean of $\{P_{ik}\}$ between
$\{T_i\}$ and $T_k$, where $i=1,2,....,k-1$,
is equal to the $P$-coefficient of
a heat pump between $T_H$ 
and $T_k$, where $T_H$ is the harmonic mean of $\{ T_i\}$.}

Alternately, we can assume individual operations of the 
machines between the hottest reservoir  $T_1$ and each of the 
lower-temperature reservoirs. 
Then we can formulate statements, equivalent to the above and
summarized below, that connect the means over the COPs 
to certain means over the lower temperatures.

($a'$) {\it The arithmetic mean of efficiencies $\{E_{1i}\}$
between $T_1$ and $\{T_i\}$,
where $i=2,....,k$ is equal to the efficiency of an engine between  
$T_1$ and the arithmetic mean of $\{T_i\}$.}

($b'$) {\it The harmonic mean
of  $\{R_{1i}\}$ between $T_1$ and $\{T_i\}$,
where $i=2,....,k$, is equal to the $R$-coefficient of
a refrigerator between $T_1$ and the harmonic mean of $\{T_i\}$. }

($c'$) {\it The harmonic mean
of $\{P_{1i}\}$ between $T_1$ and $\{T_i\}$,
where $i=2,....,k$, is equal to the $P$-coefficient of
a pump between $T_1$ and the arithmetic mean of $\{T_i\}$s.}

Now consider that for the ordered set of temperatures, 
we are given the probability  distribution $\{p_i\}$,
to connect two temperatures
$T_i$ and $T_k$ by a Carnot machine, say, a heat engine.
Then we have :
\be
\sum_i p_i E_{ik} = 1- T_k \sum_i \frac{p_i}{T_i}
\ee
Here, $\sum_i p_i /T_i = 1/\bar{T}$ defines the weighted
harmonic mean of $T_i$s.
Then the statement $(a)$ in the above, is generalized as follows:

{\it For a given probability distribution $\{p_i\}$,
the weighted arithmetic mean of 
the efficiencies $\{E_{ik}\}$ between $T_i$ and $T_k$,
is equal to the efficiency of an engine between  
$\bar{T}$ and $T_k$, where $\bar{T}$ is weighted harmonic mean
of the $\{T_i\}$,  $i=1,2,....,k-1$.}

Similar generalized statements, based on the weighted means,
can be made for the other cases mentioned above.

Finally, we consider an ordered set of $n$ positive, 
real numbers (temperatures): 
\be
T_1 > T_2 > \cdots > T_{k-1} > T_k > T_{k+1}> \cdots > T_n.  
\label{t1n}
\ee
We ask, under what condition, a temperature, say $T_k$, is one of the 
Pythagorean means? The solution generalizes the two-temperatures case, as
discussed in Section II. So, if $T_k$ is the arithmetic mean of the 
rest of the values within the set, i.e. 
$T_k = A(T_1,T_2,...T_{k-1},T_{k+1},...,T_n)$, then the following
relation can be written:

\be
\sum_{i=1}^{k-1} R^{-1}_{ik} = \sum_{i=k+1}^{n} P^{-1}_{ki}.
\label{nam}
\ee
In other words, the $R$ coefficient enters in the above, for
temperatures above $T_k$ and the $P$ coefficient, for temperatures
below $T_k$. 

Now from the definition of the harmonic mean, we have:
\be
\sum_{i=1}^{k-1} R^{-1}_{ik} = (k-1)R_{H}^{-1},
\label{rh}
\ee
and
\be
\sum_{i=k+1}^{n} P^{-1}_{ki} = (n-k)P_{H}^{-1},
\label{ph}
\ee
where $P_H$ is the harmonic mean of $\{ P_{ki} \}$.
Using Eqs. (\ref{rh}) and (\ref{ph}) in (\ref{nam}), 
we get 
\be
\frac{R_H}{P_H} = \frac{k-1}{n-k}.
\ee
It implies that the ratio of effective $R$- to $P$- coefficients
which appear as harmonic means, is the ratio of whole numbers. 
This result is somewhat reminiscent of the original observation of 
Pythagorus about the ratios of means, expressible as ratios
of whole numbers. 

Similarly, we can consider the occurence of other means within the 
given set of real, positive numbers. It is left as an 
exercise for the reader that 
if $T_k = H(T_1,T_2,...T_{k-1},T_{k+1},...,T_n)$,
then the following condition is satisfied:
\be
\sum_{i=1}^{k-1} P^{-1}_{ik} = \sum_{i=k+1}^{n} R^{-1}_{ki}.
\label{ham}
\ee
Finally, if $T_k = G(T_1,T_2,...T_{k-1},T_{k+1},...,T_n)$,
for some given $k$, then we can have three situations for  
a pair of temperatures $(T_i, T_k)$, 
with $i\ne k$: all $E$ coefficients, all 
$R$ coefficients, or all $P$ coefficients. 
Then it can be shown that
\bea
{\displaystyle \prod_{i=1}^{k-1}} (1-E_{ik}) 
&=& {\displaystyle \prod_{i=k+1}^{n}} (1-E_{ki}) \\
{\displaystyle \prod_{i=1}^{k-1}} (1+R^{-1}_{ik}) &=& 
{\displaystyle \prod_{i=k+1}^{n}} (1+ R^{-1}_{ki}),\\
  {\displaystyle \prod_{i=1}^{k-1}} (P^{-1}_{ik}-1) 
  &=& {\displaystyle \prod_{i=k+1}^{n}} (P^{-1}_{ki}-1).
\label{ngm}
\eea

\section{Summary}
Concluding, we have highlighted a few instances where
the Pythagorean means 
arise naturally in the discussion of coefficients of performance
for Carnot machines working between hot and cold reservoirs.
 In comparing the COPs of a machine 
between $T_h$ and $T_i$, and of a machine between 
$T_i$ and $T_c$, we have considered scenarios like $EE, RR, PP,
RP$, and $PR$. So far, we did not consider an engine
along with a refrigerator or a pump. The intuition behind this
was to compare the performance of similar machines. 
Thus while a refrigerator or a heat pump drives heat from
cold to hot, in a heat engine the heat flow is in 
opposite direction. Further, it is not possible to consider
a case like $EP$, because while $E<1$, we have $P>1$,
and so their COPs cannot become equal.
But scenario like $ER$ or $RE$ are possible. It can be 
shown that the equality of respective COPs
in these cases, is possible only if the condition
$\theta < 3 -2\sqrt{2}$ holds.
Then it leads to a different kind of symmetric mean, and 
not to a Pythagorean mean.

We also considered various physical situations
involving more than one reservoirs ($T_i$) 
that can be coupled via a Carnot machine to a given reservoir ($T_k$),
and the mean COP has been related to the effective COP evaluated from
the mean temperature and $T_k$. The relations pointed out in this paper and the known 
inequalities between Pythagorean means may be useful to
fix bounds on performance of thermal machines in such physical situations. 
For instance, consider the fact that the harmonic mean is dominated 
by the lowest number: $H(\{ R_{ik} \}) \le (k-1) {\rm Min}(\{ R_{ik} \})$.
From the definition, $R_{ik} = T_k / (T_i - T_k)$, we have 
${\rm Min}(\{ R_{ik} \}) = R_{1k}$, since $T_1$ is the largest entry.
Then from Eq. (\ref{rh}), we get the upper bound for the 
effective COP, $R_{Ak}  \le (k-1) R_{1k}$.

\section*{Acknowledgement}
The author is thankful to Alexander von Humboldt Foundation, 
for financial support, and to Institute for Theoretical Physics-1,
Stuttgart University, for hospitality, where a part of this
work was done. 
%

\begin{figure}
  \includegraphics[width=9.5cm]{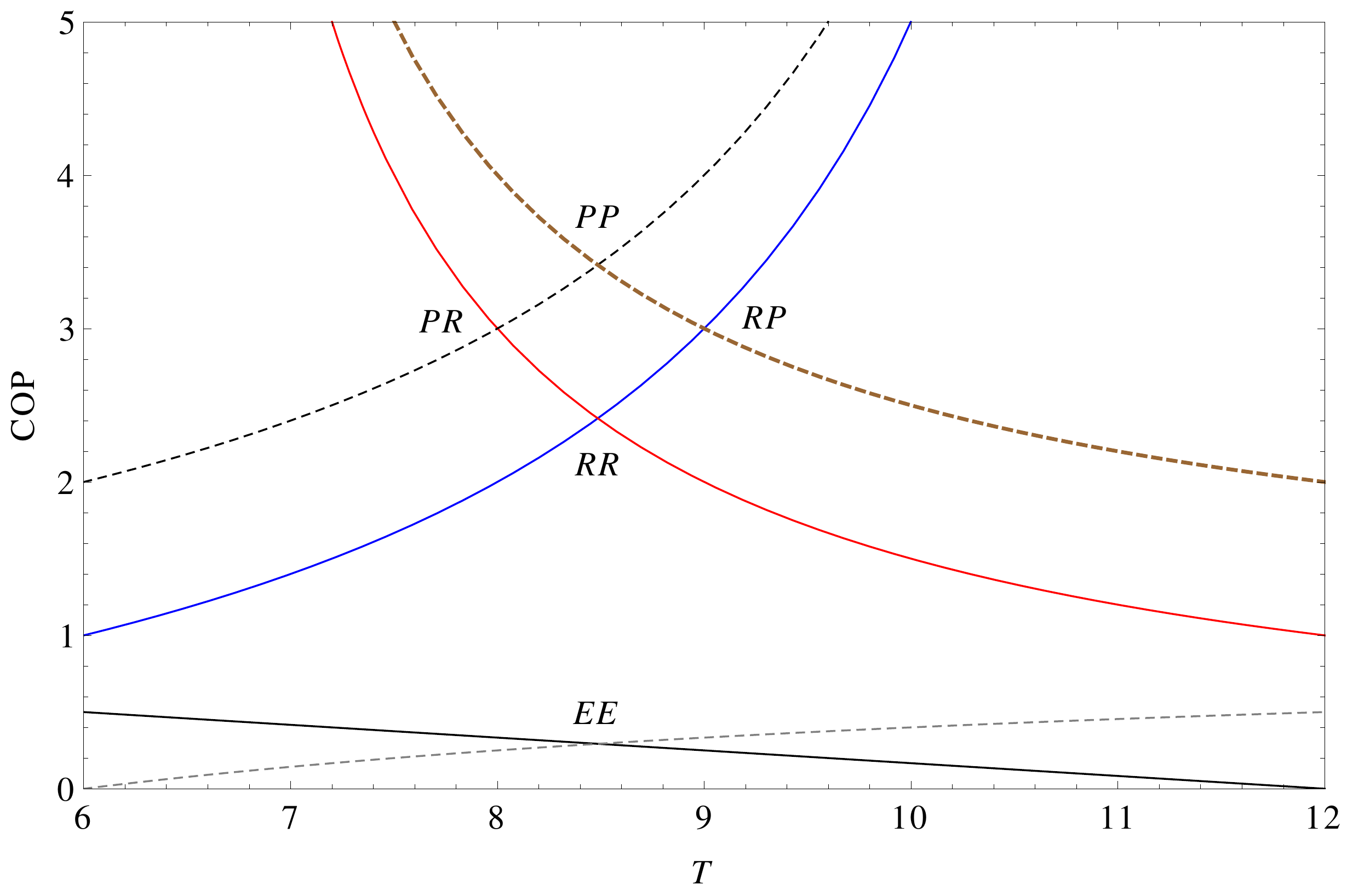}
  \caption{COPs for Carnot engines, refrigerators and pumps, 
  between one of the extreme temperatures
  $T_h=12$, or, $T_c =6$, 
  and an intermediate temperature $T$. The COPs for  $EE$ 
  (as well as $RR$ and $PP$) case intersect 
  at $T=G(T_h,T_c) = 8.485$; the intersection point for the $PR$ case is 
  at $T= H(T_h,T_c) = 8$, while that 
  for $RP$ case, is at $T = A(T_h,T_c) = 9$.}
\end{figure}

\begin{table}
 \begin{tabular}{ccc}
 \hline
 Configuration & Mean $T_i$ & \quad  COP \\
 \hline
 $EE$  & $G(T_h,T_c) $ & \quad $1-\sqrt{\theta}$   \\
 $RR$  & $G(T_h,T_c) $ & \quad $\frac{\sqrt{\theta}}{1-\sqrt{\theta}}$   \\
 $PP$   & $G(T_h,T_c)$ & \quad  $\frac{1}{1-\sqrt{\theta}}$  \\
 $PR$    & $H(T_h,T_c) $ & \quad $\frac{1+\theta}{1-\theta}$   \\
 $RP$    & $A(T_h,T_c)$  & \quad $\frac{1+\theta}{1-\theta}$  \\
 \hline  
 \end{tabular}
 \caption{Means $T_i$ where $T_h > T_i > T_c$ for various 
 configurations, as explained in the text. COPs are 
 given in terms of $\theta = T_c/T_h$.}
\end{table}

\begin{table}
 \begin{tabular}{clcc}
\hline 
 Case & \quad Machine      & \quad Mean COP & \quad Mean $\{T_i \}$ \\
 \hline
$(a)$  &\quad Engine         & $A(\{E_{ik}\})$  &\quad $H $   \\
$(b)$  &\quad Refrigerator   & $H(\{R_{ik}\})$  &\quad $A$   \\
$(c)$  &\quad Pump           & $H(\{P_{ik}\})$  &\quad $H$   \\
$(a')$ &\quad Engine         & $A(\{E_{1i}\})$  &\quad $A$    \\
$(b')$ &\quad Refrigerator   & $H(\{R_{1i}\})$  &\quad $H$    \\
$(c')$ &\quad Pump           & $H(\{P_{1i}\})$  &\quad $A$  \\
 \hline  
 \end{tabular}
 \caption{Means in the presence of multiple reservoirs
 with temperatures $T_1 > T_2 > \cdots > T_{k-1} > T_k$.
 The table shows type of the mean of COPs versus the mean
 of the reservoir temperatures $\{T_i\}$, where $i=1,2,...,k-1$ for 
 cases $(a), (b)$ and $(c)$, while $i=2,3,...,k$ for cases $(a'), (b')$ and $(c')$.}
\end{table}

\fbox{\begin{minipage}{35em}
      \bc
      BOX 1 \\
      {\bf The Music in the Means} 
      \ec
      In ancient Greece, like during the days of Pythagorus 
      (circa 6th century B.C), the mean (of two numbers) 
      represented a kind of balance between 
      the two opposites of high and low. 
      The three means $A,G,$ and $H$ were regarded
       as forming a triad and considered to be fundamental.
      The followers of the Pythagorean 
      school were enthused all the more to find these means
      reflected in the order and harmony of nature. 
According to legend, one day, as Pythagorus was walking past
a blacksmith's shop, he heard a harmonious ringing of hammers.
On closer observation, he found that when the hammers with
masses in a specific proportion, were sounded,
the sound appeared musical to human ears. 
Fascinatingly, he found these proportions to be given in terms of
whole numbers, such as $12: 9 = 8 : 6$. 
Here 9 is the arithmetic mean of the extreme numbers 12 and 6.
Similarly, 8 is the harmonic mean of 12 and 6. In general,
for two numbers $a > b$, it implies the ratio
      \be
       a: \frac{a+b}{2} = \frac{2ab}{a+b}: b.
\nonumber
       \ee
       These proportions form the basis for tuning theory
       in music and represent the ratios of frequencies. 
       For example, here 6 and 12 represent the fundamental 
       and the octave. Then the arithmetic mean frequency 
       gives the perfect fifth, and the harmonic mean frequency gives 
        the perfect fourth. Other intervals on musical 
        scale can also be characterized in a similar way. 
              
      \end{minipage}}

      \fbox{\begin{minipage}{35em}
      \bc
      BOX 2 \\
      {\bf Means and Thermodynamics} 
      \ec
   There are
   various proofs, including geometric ones,  of
   simple inequalities between the means,
   such as the arithmetic
   mean-geometric mean inequality: $A\ge G$. 
   Thermodynamic models are especially useful in 
   giving a kind of 'physical' proofs for these 
   inequalities. Thus consider $n$ systems 
   with a constant heat capacity
$C$ and initial temperatures, 
$\{ T_i |i=1,...,n \}$. Upon mutual
thermal contact, these systems arrive at equilibrium
with a common final temperature, say $T_f$. 
From the first law of thermodynamics, the initial
and final energies are equal, so that
\[ 
\sum_{i=1}^{n} C(T_i - T_f) = 0,
\]
which implies $T_f = \sum_i T_i /n$.
The total entropy change during the process is:
\bea
\Delta S &=&  \sum_{i=1}^{n} \int_{T_i}^{T_f} \frac{C}{T}\;dT  \nonumber \\
  &=& n C \left[ \ln T_f - \ln (\Pi_i T_i)^{1/n} \right].
  \nonumber
\eea
  So by virtue of the $A$-$G$ 
inequality, we get $\Delta S > 0$. Arnold
Sommerfeld  posed this problem in his Lectures on 
Theoretical Physics Vol. 5, as an application of the second
law to prove an algebraic inequality, i.e.,
if we assume the validity of the macroscopic form 
of second law ($\Delta S >0$), then the $A$-$G$ inequality follows. 
P. T. Landsberg also endorsed the viewpoint that herein the truth
of pure mathematics is directly accesssible from the principles
of science. It is understandable if some authors regard  mathematical 
inequalities to be more fundamental, 
something that exists prior to scientific concepts.  
At best, the thermodynamic or physical models may provide
a kind of heuristic which makes these inequalities seem plausible.
      \end{minipage}}

      \end{document}